\documentclass[useAMS,usenatbib,aps,floatdix,nofootinbib]{mn2e}

\usepackage{color}

\usepackage{graphics}
\usepackage{amssymb,amsmath}
\usepackage{amsmath}
\usepackage{psfrag}
\usepackage{graphicx}
\newcommand{\ve}[1]{\mathbf{q1}}
\newcommand{\f}{\frac}

\newcommand{\be}{\begin{equation}}      
\newcommand{\ee}{\end{equation}}      
      
\newcommand{\bef}{\begin{figure}}      
\newcommand{\eef}{\end{figure}}      
\newcommand{\bea}{\begin{eqnarray}}    
\newcommand{\eea}{\end{eqnarray}}      

\newcommand{\av}[1]{\ensuremath{\left\langle q1 \right\rangle}}
\def\gtapprox{\mathrel{\spose{\lower 3pt\hbox{$\mathchar"218$}}      
\raise 2.0pt\hbox{$\mathchar"13E$}}}      

\def\sposeq1{\hbox to 0pt{q1\hss}}
\def\ltapprox{\mathrel{\spose{\lower 3pt\hbox{$\mathchar"218$}}
\raise 2.0pt\hbox{$\mathchar"13C$}}}
\def\gtapprox{\mathrel{\spose{\lower 3pt\hbox{$\mathchar"218$}}
\raise 2.0pt\hbox{$\mathchar"13E$}}}
\def\inapprox{\mathrel{\spose{\lower 3pt\hbox{$\mathchar"218$}}
\raise 2.0pt\hbox{$\mathchar"232$}}}

\newcommand{\tve}[1]{\tilde{\boldsymbol{q1}}}

\def\bse{\begin{subequations}}
\def\ese{\end{subequations}}

\def\lsim{\raise 0.4ex\hbox{$<$}\kern -0.8em\lower 0.62ex\hbox{$\sim$}} 
\def\gsim{\raise 0.4ex\hbox{$>$}\kern -0.7em\lower 0.62ex\hbox{$\sim$}}

\def\f0N{f_0^{(N)}}
\def\bec{\begin{center}}
\def\eec{\end{center}}

\title[Particle number dependence in $N$-body systems] 
{Particle number dependence in the non-linear evolution 
of $N$-body self-gravitating systems} 
\author[Benhaiem et al.]
{D. Benhaiem${^{1,2}}$, M. Joyce${^{2}}$, F. Sylos Labini${^{3,1,4}}$
  and T. Worrakitpoonpon${^{5}}$
  \\
  $^{1}$Istituto dei Sistemi Complessi Consiglio Nazionale delle Ricerche, Via dei Taurini 19, 00185 Rome,   Italy
  \\
  $^{2}$Laboratoire de Physique Nucl\'eaire et de Hautes \'Energies, UPMC IN2P3 CNRS UMR 7585, Sorbonne Universit\'es,
  \\ 4, place Jussieu, 75252 Paris Cedex 05, France
  \\
  $^{3}$Centro Fermi - Museo Storico della Fisica e Centro Studi e Ricerche “Enrico Fermi'', 00186 Rome, Italy
  \\
  $^{4}$INFN Unit Rome 1, Dipartimento di Fisica, Universit\'a di Roma Sapienza, Piazzale
  Aldo Moro 2, 00185 Roma, Italy
  \\
  $^{5}$Faculty of Science and
  Technology, Rajamangala University of Technology Suvarnabhumi,
  Nonthaburi Campus, \\ Nonthaburi 11000, Thailand}

\begin{document}

\date{\today}

\maketitle

\begin{abstract}

Simulations of purely self-gravitating $N$-body systems are often 
used in astrophysics and cosmology to study the collisionless limit 
of such systems. Their results for macroscopic quantities should 
then converge well for sufficiently large $N$. Using a study of the evolution 
from a simple space of spherical initial conditions --- including a region
characterised by so-called ``radial orbit instability'' --- we
illustrate that the values of $N$ at which such convergence is
obtained can vary enormously. In the family of initial conditions we
study, good convergence can be obtained up to a few dynamical times
with $N \sim 10^3$ --- just large enough to suppress two body
relaxation --- for certain initial conditions, while in other cases
such convergence is not attained at this time even in our largest
simulations with $N \sim 10^5$. The qualitative difference is due to
the stability properties of fluctuations introduced by the $N$-body
discretisation, of which the initial amplitude depends on $N$. 
We discuss briefly why the crucial role which such fluctuations 
can potentially play in the evolution of the $N$ body system
could, in particular, constitute a serious problem in 
cosmological simulations of dark matter. 
\end{abstract}


\section{Introduction} 

A purely Newtonian self-gravitating system of a large (but finite)
number $N$ of identical particles is a fundamental idealized model in
astrophysics and cosmology.  Usually the aim of numerical simulations
of such systems is to reproduce the collisionless limit.  In this
limit the system's evolution is described by the collisionless
Boltzmann equation for the smooth phase space density. The latter
corresponds to taking $N \rightarrow \infty$ at fixed time with the
phase space mass density converging to a continuous function. The $N$
bodies of the simulation are then not physical entities (e.g dark
matter particles or stars) but numerical ``macro-particles". Any
dependence on their number (or other properties) is an artefact of the
use of the $N$ body method employed, or a ``discreteness
effect"\footnote{We note, however, that there have recently been
  studies of the direct solution of the collisions Boltzmann equation
  starting from cosmological type initial conditions
  \citep{yoshikawa_etal_2013} and also certain spherically symmetric
  initial conditions \citep{colombi_etal_2015,
    colombi+sousbie_2015}.}.  To obtain a precise result for the
collisionless limit, the particle configuration must sample faithfully
the continuum phase space density. By construction this condition is
respected at the initial time, up to quantifiable finite $N$ sampling
fluctuations. It is difficult, however, to determine a priori how well
it is respected during the system's evolution: given the highly
complex non-linear dynamics, differences between the evolution of the
finite $N$ sampling, and that of the continuum limit may be very
non-trivial. In practice this issue is addressed by studying,
numerically, the dependence of results on $N$, to establish as far as
possible whether the relevant results do indeed converge to $N$
independent values.  The question of {\it how many particles are in
  practice needed to simulate accurately the collisionless limit (up
  to some specified time)} can only be answered in the context of a
given problem, and indeed different estimates can be found in the
literature in different contexts (e.g. \cite{roy+perez_2004,
  elzant_2006, diemand, discreteness2_mjbm, sellwood_2015}).  For
example if finite $N$ effects are dominated by two body collisions,
such effects are negligible provided one limits the evolution to a
time scale short compared to the characteristic time for such effects,
of order $\sim (N/\log N) \,\tau_{dyn}$, where $\tau_{dyn}$ is a
characteristic time of the collisionless (mean field) dynamics (see
e.g. \cite{binney}).

In this paper we address this issue through the study of the evolution
a simple class of initial conditions --- initially spherical density
profiles --- using $N$ body simulations.  The study of this specific
case allows us to illustrate clearly how fluctuations introduced by
the finite $N$ discretisation intrinsic to $N$ body simulations {\it
  may}, in the presence of instabilities in its non-linear evolution,
play a crucial role in the macroscopic dynamics of the system on
timescales which are very much shorter than those typical of
collisional effects.  Furthermore the $N$ dependence of physical
quantities can be extremely weak and difficult to detect in a
numerical convergence study unless one knows precisely where to look
for them as we do in our simple chosen system.

It is perhaps useful for the reader that we recall the generic phases
of the evolution of isolated systems of a large number $N$ of
self-gravitating particles, and place our study in this broader
context. The evolution of such a system is believed to be
characterized by two distinct time-scales: the {\it dynamical time}
$\tau_{dyn}$, which characterizes the mean-field or collisionless
dynamics described by the collisionless Boltzmann (or Vlasov-Poisson)
equation, and a much longer {\it collisional} time-scale
characterizing the dynamics beyond this mean-field limit. The ratio of
the latter to the former time-scale diverges with $N$, and in the case
that the collisional dynamics is dominated by two body collisions the
relative scaling is predicted to be given, as mentioned above, by the
ratio $N/\log N$ (for an unsmoothed gravitational potential).  Thus,
starting from a generic initial condition, such a system undergoes an
evolution in (at least) two distinct phases: (1) collisionless
evolution which is observed to lead it on time scales of order
$\tau_{dyn}$ to a virial equilibrium which is a stationary state of
the collisionless Boltzmann equation. This is the process of
``collisionless" (or ``mean-field") relaxation, involving in general
phase mixing and so-called ``violent" relaxation; (2) on much longer
timescales, of order $N\,\tau_{dyn}$, such a state evolves
continuously while remaining at virial equilibrium, in principle
towards states of ever increasing entropy. This second phase of
``collisional relaxation" is more challenging to study numerically
(because of the need to simulate accurately the collisional
processes), and is the subject of many studies (see e.g.
\cite{theis+spurzem_1999, gabrielli_etal_2010, sellwood_2015}, and
references therein).

In the present work we aim to study only the first phase, of
collisionless evolution. In our numerical simulations this corresponds
to using a sufficiently large number of particles $N$ so that the time
scale of collisional relaxation is very long compared to those we
consider. {\it The dependence on $N$ of the evolution we discuss has
  therefore nothing to do with collisional effects, but arises from
  the dependence of the collisionless evolution on the particle number
  $N$ arising from how the initial conditions are sampled}.  We focus
on the case that the collisionless relaxation may itself occur in two
distinct phases: initial relaxation to a virial equilibrium which is a
stationary, but unstable, state of the collisionless Boltzmann
equation, and a second phase in which the system relaxes to a stable
stationary state.  Such a two stage dynamics can occur in particular
in the case of so-called ``radial orbit instability", an instability
of stationary states of the collisional Boltzmann equation with
spherical symmetry and purely radial orbits originally discussed by
\cite{antonov1973, henon_1973} (for a review and a more complete set
of references see e.g. \cite{marechal+perez_2011}). As shown first
numerically in \cite{merritt+aguilar_1985}, and subsequently by many
other studies, such an evolution occurs when the N-body system starts
from certain initial conditions which are cold (i.e. are strongly
sub-virial) and very close to spherical symmetry.  We choose to study
this particular kind of case here because it turns out that the
presence of the instability in the dynamics is associated with a
subtle $N$-dependence of the evolution.

The paper is organized as follows. We first present the details of the
numerical simulations and initial conditions in Sect.\ref{numsim}.  In
the following section we present the essential results of our
convergence study. We conclude with a discussion of these results,
which we also compare with some previous ones in the literature, in
particular with the study of \cite{boily+athanassoula_2006} which has
drawn attention previously to the non-trivial evolution with particle
number in numerical experiments similar to the ones we consider.  In
our conclusions we discuss in particular the possible implications of
our results for the non-linear regime of cosmological N-body
simulations of structure formation in the universe.

%%%%%%%%%%%%%%%%%%%%%%%%%%%%%%%%%%%%%%%%%%%%%%%%%%%
\section{Numerical simulations} 
\label{numsim}

\subsection{Initial conditions} 

The initial conditions of which we study the evolution are $N$
particle configurations with (i) positions sampled randomly,
in two different ways detailed below, to follow an average radial 
density decay $\rho (r) \propto r^{-1}$ away from a centre
(where $r$ is the radial distance from the latter), and 
(ii) velocities sampled randomly (and independently for
each particle) from a distribution which is uniform
and non-zero inside a sphere in velocity space
(and thus isotropic).  The radius of this sphere can be 
characterized conveniently by the initial virial ratio,
which we define as $b_0=-2K_0/W_0$ where $K_0$
is the initial kinetic energy and $W_0$ the initial potential energy
of the configuration. 

We choose this particular spatial profile as it is a simple one which is known
to give rise to qualitative evolution which depends strongly on $b_0$.
More specifically it is known to manifest strong breaking of spherical
symmetry for sufficiently low initial virial ratios (see
e.g. \cite{merritt+aguilar_1985, aguilar+merritt_1990,
  barnes_etal_2009, SylosLabini+Benhaiem+Joyce_2015}).

Because it is the dependence on $N$ of results we focus on in this paper,
and in particular how such a dependence can arise via the finite $N$
fluctuations to the average space and velocity distributions, it is of
potential importance to know how precisely the finite $N$
configurations are generated. We have considered the following ways of
generating the particle realisations:

\begin{itemize}

\item $N$ particles are distributed randomly with uniform
  density inside a sphere, and the radii of the particles are then
  rescaled to give the desired radial density decay.  The fluctuations
  in any volume are thus Poissonian in their dependence on the
  particle number.
 
\item A sphere containing $N$ particles is extracted from
a distribution of particles initially located on a regular (simple cubic)
  lattice, the sphere being centred on the centre of a lattice cell.
  The radial positions of the particles are then ``stretched'' to give the
  desired radial density decay. In this case both the radial and
  angular fluctuations are very suppressed compared to those of the
  configurations obtained by the first method.

\end{itemize}

As a consequence of the different spatial sampling, the
velocity fluctuations are also slightly different in the two cases.

In all cases the limit $N \rightarrow \infty$ of the algorithm
converges to the continuum phase space density, and the amplitude of
the fluctuations around this density is a monotonically decreasing
function of $N$. As we will discuss in our conclusions an
extrapolation $N \rightarrow \infty$ of these initial conditions 
obtained as some finite $N$ could be performed in different 
manners leading, in certain cases, to different results.

\subsection{Quantities studied } 

In our complete study we have monitored the evolution, starting from
many realisations of the above class initial conditions, of numerous
macroscopic quantities: kinetic and potential energy, bound mass,
angular momentum of bound and ejected mass, parameters charactering
the shape of the structure, density profiles in spherical shells,
velocity distributions etc.. We report here results for only one
chosen macroscopic parameter. The study of this parameter suffices to
explain our central finding about the dependence on $N$ of the
evolution of macroscopic quantities. This quantity is the shape
parameter charactering the deformation from sphericity of the
structure, and defined as
\be 
\label{iota} 
\iota_X(t)= \frac{\Lambda_1(t)}{\Lambda_3(t)} -1 \;.  \ee 
where $\Lambda_1 \ge \Lambda_2 \ge \Lambda_3$ are the eigenvalues of
the inertia tensor of the $X$ percent most bound particles (i.e. of
the first $X$ percent of the bound particles when they are ranked by
increasing energy).  We will consider $X=80$, which typically
characterizes well the final global shape of most of the mass.

\subsection{Numerical simulations} 

We have used the N-body code {\tt Gadget2} \citep{springel_2005}.  All
results presented here are for simulations in which energy is
conserved to within {\it one tenth of a percent} up to the longest 
times evolved to. In order to achieve such precision, even for the very 
violent collapses which result from the colder initial conditions, we 
have employed an accuracy of time integration significantly greater than 
that typically used for the code for simulations of this kind~\footnote{The time-stepping parameter 
$\eta$ has been set to $\eta=0.005$, compared to $\eta=0.025$ as typically used.}.  
Further the parameters controlling the accuracy of the force calculation have been 
set so that the force is in practice calculated {it by direct $N^2$ summation}. 
For the simulations reported here, the force softening parameter 
has been taken to be $\epsilon = (0.056)RN^{-1/3}$ where $R$ is the 
initial radius of the sphere, i.e., we use a softening which is fixed in 
proportion to the mean interparticle separation. We have performed extensive 
tests (see also \cite{Benhaiem+SylosLabini_2015,SylosLabini+Benhaiem+Joyce_2015} 
for more details) of the effect of varying the force smoothing parameter 
in a wide range. In particular we have verified for several cases that we 
obtain fully consistent  results taking  $\epsilon/R=10^{-3}$.  In 
general we obtain very stable results provided it is significantly smaller 
than the minimal size reached by the whole structure during collapse. 
We note that the smoothing given above  is also consistent with a value
always below the maximal one recommended by \cite{boily+athanassoula_2006} 
in their study of the convergence of simulations of this kind. 

\section{Results} 
\label{results}

As unit of time we take the characteristic time for the
mean-field (collisionless) evolution,
\be 
\tau_{dyn}=\sqrt{ \frac{3 \pi}{32 G \rho_0} } \,.
\ee
where $\rho_0$ is the initial mean mass density of the sphere (and $G$
the Newton's constant).  The systems we consider starting from the
above classes of initial conditions are in all cases well virialized
(i.e., with only very small fluctuations of the viral ratio) by $t
\approx 2$, and we run them typically up to a time roughly an order of
magnitude longer.

Results for evolution from these initial conditions for various $b_0$
and $N$ have been reported in various previous studies (e.g.
\cite{merritt+aguilar_1985, aguilar+merritt_1990, barnes_etal_2009}).
We have checked that our results, when comparable, appears to be very
consistent, quantitatively and qualitatively, with previously published
ones. In particular, as we will see below, we observe, for $b_0 \lsim
0.1 $ a virialized state which strongly breaks spherical
symmetry~\footnote{One minor difference is that most authors have
  employed a gaussian distribution for the initial velocities, rather
  the uniform one we use. This is not expected to, and indeed does not
  appear to, make any significant difference in the context of this
  study.}.

\subsection{N dependence of ``final" quantities}
\label{Ndependence-final}
\begin{figure}
\vspace{1cm}
{
\par\centering \resizebox*{9cm}{8cm}{\includegraphics*{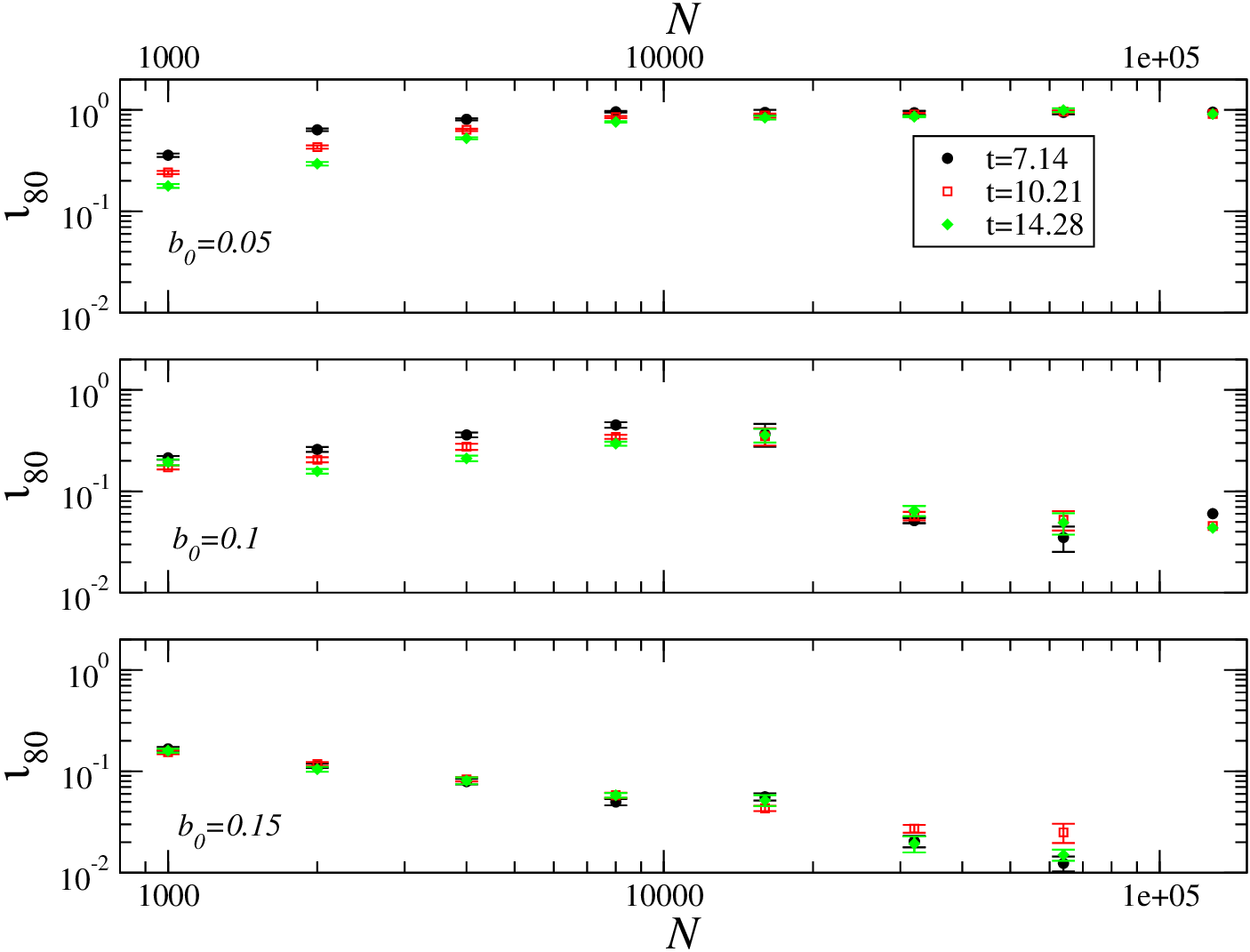}}
\par\centering
}
\caption{Each panel shows $\iota_{80}$ for a given $b_0$ as a function
  of $N$.  In each case its value, averaged over an ensemble of
  realisations of Poissonian initial conditions (see footnote in text 
  for details of number of realizations), is plotted at three
 different times, well after the virialisation of the system (at $t  \approx 2$).}
\label{FIG_I80_N} 
\end{figure}

We have studied the evolution for many values of $b_0$ in the range
between $0$ and $0.3$.  Shown in Fig.~ \ref{FIG_I80_N} are results for
the measured $\iota_{80}$ at three times as a function of $N$ 
three chosen values of the initial viral ratio $b_0$ ($b_0=0.05$,
$b_0=0.1$ and $b_0=0.15$) and for Poissonian initial
conditions. Each point is the average value calculated
over $M$ of realisations of the given initial condition~\footnote{The
  number $M$ of realisations are as follows: $M=70$ for $N=10^3$,
  $M=40$ for $N=2 \times 10^3$, $M=30$ for $N=4 \times 10^3$ and
  $N=8\times 10^3$, $M=6$ for $N=1.6\times 10^4$, $M=4$ for
  $N=3.2\times 10^4$, $M=2$ for $N=6.4\times 10^4$, and $M=1$ for
  $N=1.28\times 10^5$.}. The error bars indicate the estimated error
on the mean.

As noted the system contracts to its minimal size at $t \approx 1$ and
then re-expands and is already well virialized by $t \approx 2$. The
results for the different times indicated (starting from $t \approx
7$) and plotted in Fig.~\ref{FIG_I80_N} thus monitor the stability of
the measured shape parameter $\iota_{80}$ at ``long times". A very
strong stability of the result is in fact observed in all but the
simulations at the lowest $N$ --- with up to a couple of thousand
particles --- in the case $b_0=0.05$. Such an evolution at lower
particle number --- towards a more spherically symmetric structure ---
is, as has been studied in detail by \cite{theis+spurzem_1999} for
similar initial conditions --- due to two body collisionality. Beyond
$N=4000$ such effects do not appear to play any role in the evolution
on the time scales we simulate, and the values of $\iota_{80}$
obtained at a given $N$ are very stable in time.

Let us consider now the dependence on $N$ of these ``final''
$\iota_{80}$. Depending on the value of $b_0$, the plots show very
different qualitative behaviours: for $b_0=0.05$ (upper panel) an
apparently asymptotic value of $\iota_{80} \approx 1$ is reached at $N
\approx 10^4$; for $b_0=0.15$ (lower panel) $\iota_{80}$ is always
much smaller and appears to decrease monotonically to zero as $N$
increases to its largest value. We will see below that the value of
$\iota_{80}$, even in the cases where it is very small, is always
measurably larger than the initial value of $\iota_{80}$, i.e.  the
measured asymmetry, albeit small, is above the level of the intrinsic
noise introduced by sampling the density field with a finite $N$.  For
the intermediate case, $b_0=0.1$ (middle panel), we see a very
different behaviour, which is, very roughly, an interpolation, in a
narrow range of $N$ just above $N=10^4$, between the other two cases:
up to $N=16000$ the final $\iota_{80}$ grows up to an average value of
$\iota_{80} \approx 0.4$, while at $N=32000$ it is about an order of
magnitude smaller. In summary, in the first and last case the
behaviours as a function of $N$ are apparently a simple monotonic
convergence towards an asymptotic value which is already very well
approximated for a few thousand particles. In the intermediate case
such a convergence towards the asymptotic value is observed only when
there is roughly an order of magnitude more particles.
  
What lies behind these very different behaviours, both quantitative
and qualitative, are the physical differences between the non-linear
dynamics in the different cases: for $b_0=0.05$ the system strongly
breaks spherical symmetry, while at $b_0=0.15$ this is not the case.
As discussed in the literature (see references above), this symmetry
breaking is in principle due to so-called ``radial orbit instability":
stationary spherically symmetric solutions of the collisionless
Boltzmann equation are unstable when the velocity dispersion is
purely, or strongly, radial as originally shown by
\cite{Polyachenko_1981}. Such an instability exists for some range of
the parameter (or parameters) characterising the anisotropy of the
velocity distribution, and is expected to correspond to an existence
of the instability starting out of equilibrium at values of $b_0$ up
to some critical value. The simplest interpretation of the behaviour in
the case $b_0=0.1$ is then that this value is just above the critical
value, but sufficiently close to it that the larger intrinsic
fluctuations about spherical symmetry present at smaller $N$ can allow
the system to access the instability. This behaviour can be 
considered to be an example of the sensitivity to finite size effects
characteristic of critical phenomena.  
 
The naive interpretation of these results is thus
the following. Convergence (to within some chosen precision)
to the continuum result is obtained in all cases
with a sufficiently large $N$. This number may, however,
be very significantly larger if the initial condition is close to a 
``critical" point where the final state obtained
changes qualitatively. Unless one is very close
to such a point, on the other hand,  it appears that 
the only finite $N$ effect in a simulation one needs 
to worry about in this system is two body collisionality. 
We will see now that these conclusions are 
not valid when we study the temporal evolution carefully.

\subsection{$N$ dependence of temporal evolution} 

\begin{figure}
\vspace{1cm}
{
\par\centering \resizebox*{9cm}{8cm}{\includegraphics*{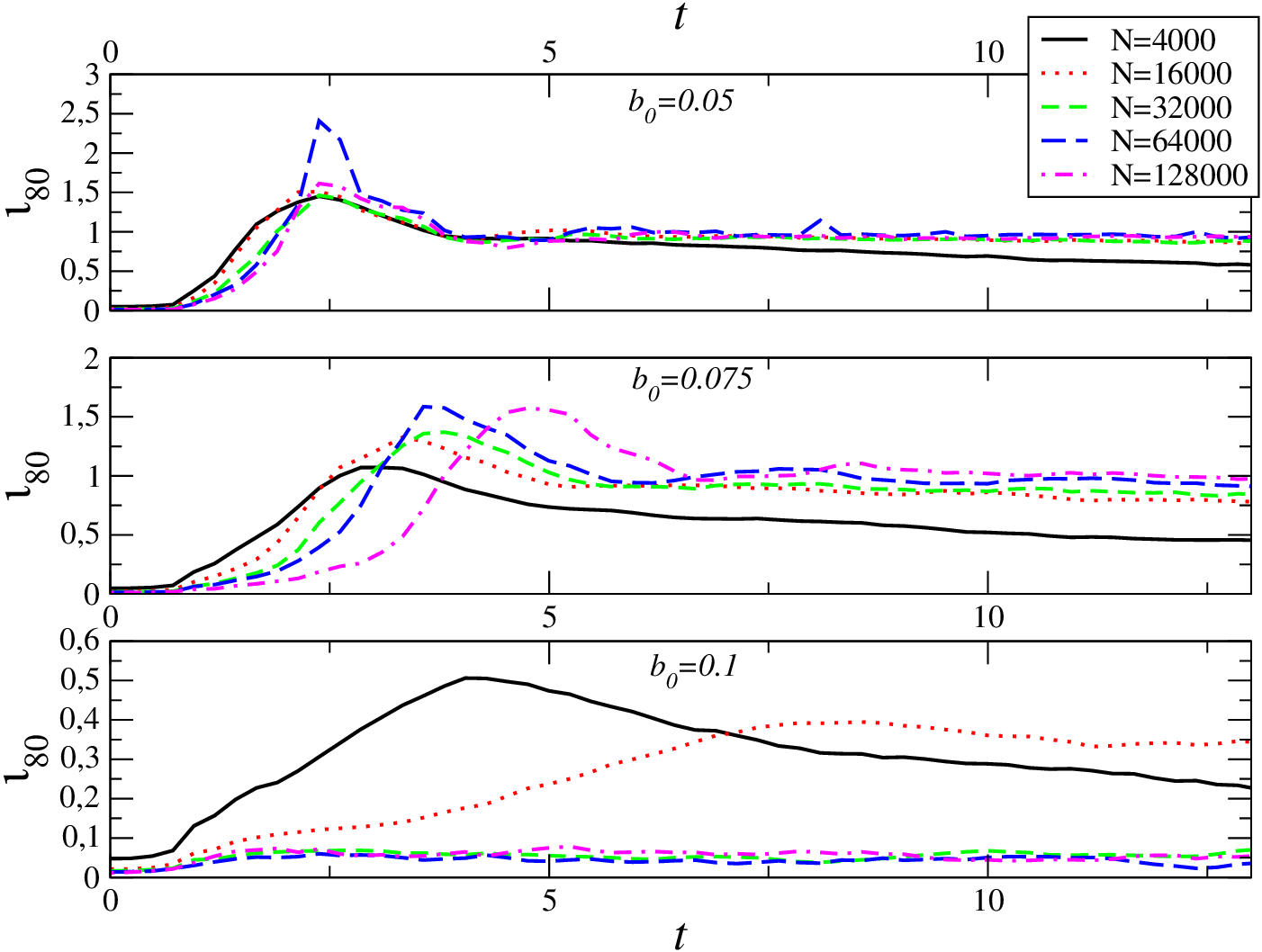}}
\par\centering
}
\caption{Each panel shows, for a given fixed value of $b_0$, the 
temporal evolution of $\iota_{80}$ as a function of time averaged
over an ensemble of realisations with the indicated values of
$N$. (The number of realisations in each case is the same 
as in the previous figure).}
\label{FIG_I80_T} 
\end{figure}

The above analysis relies on the study of the convergence of the
macroscopic properties at relatively long times, well after the system
has ``settled down" to virial equilibrium.  If we are only interested
in the state of the system at asymptotically long times this is
sufficient. However if, as for example in the context of cosmological
simulations, the temporal evolution of macroscopic quantities needs to
be resolved, convergence tests must be applied to these. Shown in
Fig.~\ref{FIG_I80_T} is the evolution of $\iota_{80}$ as a function of
time, starting from realisations of our Poissonian initial conditions
with the indicated values of $b_0$, for a range of different values of
$N$. If we limit ourselves to examining the curves for $t > 7$, our
conclusions about convergence are those obtained above from
Fig.~\ref{FIG_I80_N}: for a sufficiently large $N$ there appears to be
clear convergence of the curves. However, considering now also times
$t < 7$ for the cases $b_0=0.05$ and $b_0=0.075$ (i.e. in which the
final state breaks radial symmetry), the convergence is much poorer,
or even apparently not attained at all: in particular the time at
which the instability in $\iota_{80}$ develops in the case $b_0=0.075$
visibly depends on the number of particles, with the time at which
$\iota_{80}$ attains its maximum shifted {\it by almost two dynamical
times} between the largest two simulations. In the case $b_0=0.05$ the
difference in this time is much smaller, but the details of the
evolution around the maximum appears to be quite different. In either
case it is clear that even at $N=128000$ one cannot conclude that the
temporal macroscopic evolution has converged.

\begin{figure}
\vspace{1cm}
{
\par\centering \resizebox*{9cm}{8cm}{\includegraphics*{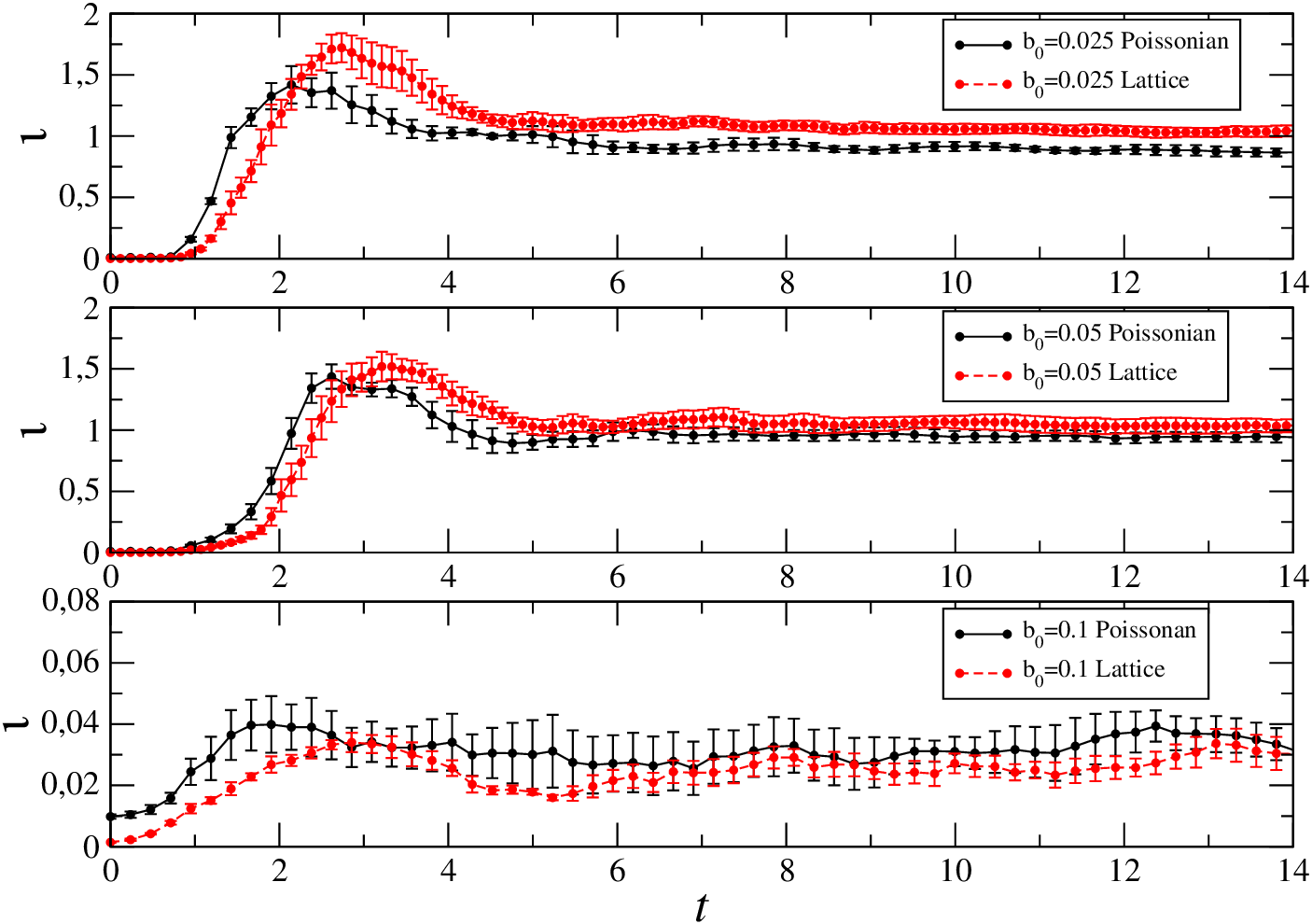}}
\par\centering
}
\caption{Each panel shows, for a given value of $b_0$, the 
mean temporal evolution of $\iota_{80}$, and its estimated
error, 
in five realisations with $N=10^5$ particles of Poissonian (filled lines) and 
lattice (dashed line) initial conditions. }
\label{FIG_I80_POI_LAT} 
\end{figure}

\begin{figure}
\vspace{1cm}
{
\par\centering \resizebox*{9cm}{8cm}{\includegraphics*{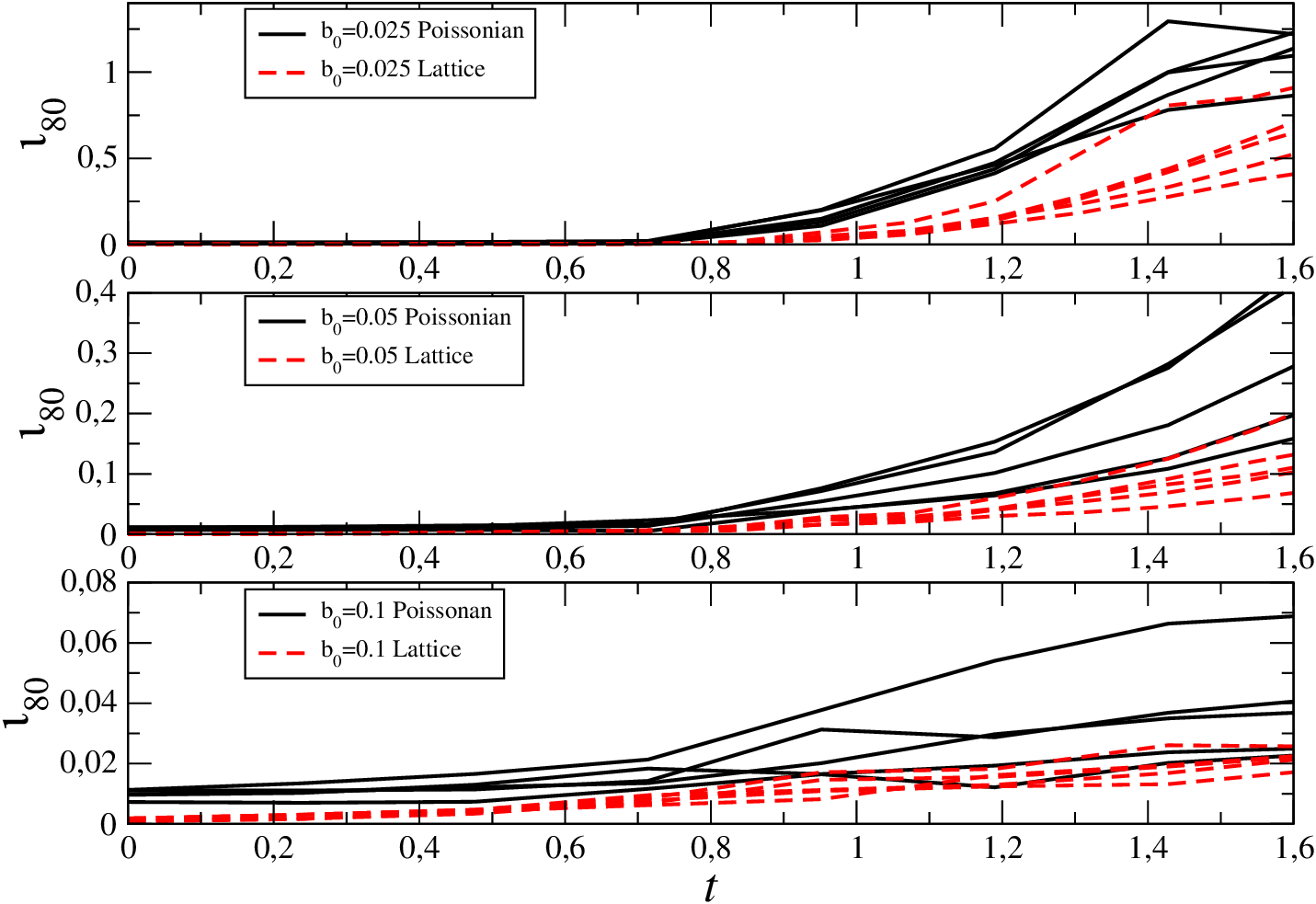}}
\par\centering
}
\caption{Evolution at early times ($0.5 < t < 1.5$)  of $\iota_{80}$ 
in the five realisations of Poissonian (filled lines) and 
lattice (dashed line) initial conditions from which the averages in the
previous figure are calculated.}
\label{FIG_I80_POI_LAT_REALS} 
\end{figure}

The shift of the time at which the instability in $\iota_{80}$
develops is explained simply making the hypothesis that the effective
seeds for its development are the finite $N$ fluctuations in the
initial conditions which break the spherical symmetry: as $N$
increases their amplitude is decreased, and thus their development
retarded. This can be further tested by comparing evolution at given
$N$ from our two different sets of initial conditions, Poissonian and
lattice-like. Given that, at fixed $N$, the amplitude of the
density fluctuations around perfect spherical symmetry are
suppressed in the lattice initial conditions compared to those in
the Poissonian initial conditions, the seeds for the instability are
smaller. Thus, if our interpretation of the previous results is
correct, we would anticipate that there should be a delay in the
development of the instability at a given level.  Shown in
Fig.~\ref{FIG_I80_POI_LAT} are plots of the evolution of $\iota_{80}$
averaged over five realisations of the two cases, for two values of
$b_0$ in the range where there is symmetry breaking, and also for the
case $b_0=0.1$. In the first two cases, including the one
corresponding to the upper panel in Fig. ~\ref{FIG_I80_T} in which the
$N$ dependence of the temporal evolution was not so evident in this
figure, we indeed see a clear delay in the onset of the
instability for the lattice initial conditions. This
is confirmed by Fig.~\ref{FIG_I80_POI_LAT_REALS} which
shows the evolution at early times, in the time window
$[0.5, 1.5]$, for the individual realisations in each case.
The lower panel in both figures show that, for the case $b_0=0.1$ 
where the  final symmetry is very small, there is also
a measurable relative delay of the development
of the asymmetry which depends on the initial amplitude.
As the evolution continues (see Fig.~\ref{FIG_I80_POI_LAT}), 
however, the difference between the realisations appears to be 
more or less wiped out,  in contrast in particular to the case 
$b_0=0.025$, in which  the memory  of the very subtle difference 
in the initial conditions appears clearly to persist up to the
 longest time simulated.

\begin{figure}
\vspace{1cm}
{
\par\centering \resizebox*{9cm}{8cm}{\includegraphics*{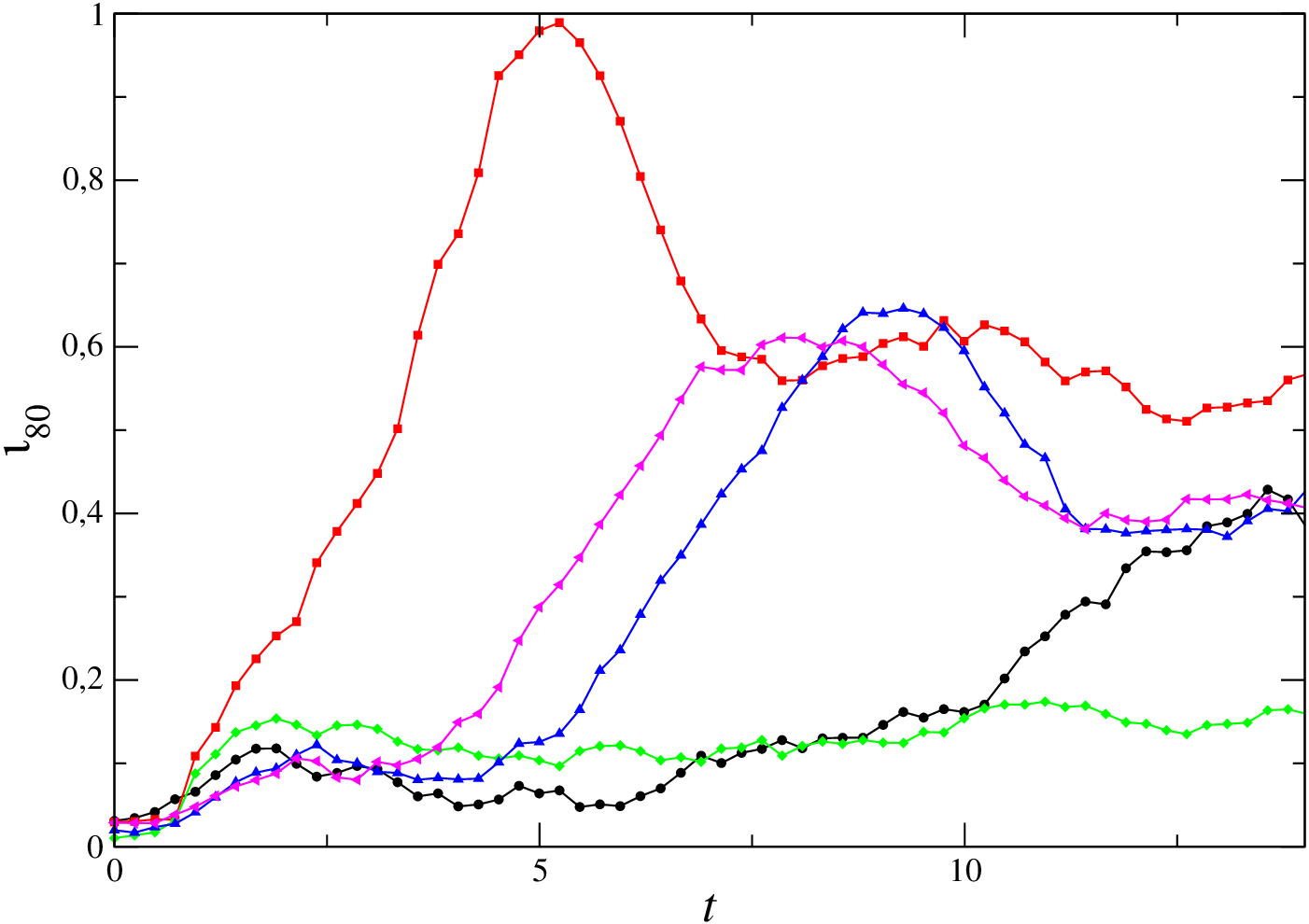}}
\par\centering
}
\caption{Temporal evolution of $\iota_{80}$ in different realisations with 
$N=16000$ of initial conditions with $b_0=0.1$.
}
\label{FIG_I80_n16000_B01} 
\end{figure}

The data we have shown up to now is averaged over a number of
realisations of each initial condition.  For most cases the dispersion
between the single realisations is small once $N$ is greater than a
few thousand. We have observed, however, that this is not always true:
in cases where the evolution is very sensitive to the value of $N$,
e.g., for $b_0=0.1$ around the transition observed between $N=10^4$
and $N=10^5$, there is also a very large dispersion between the
different realisations.  This is illustrated in
Fig. ~\ref{FIG_I80_n16000_B01} which shows the evolution in different
realisations of the $b_0=0.1$ Poissonian initial condition with
$N=16000$ particles. Remarkably we find that the macroscopic evolution
can be completely different from realisation to realisation, i.e. the
system manifests an apparent chaoticity of its {\it macroscopic}
behaviour.  We note that behaviour of this kind (``macroscopic
stochasticity'') has been observed previously in N-body systems
modelling galaxies by \cite{sellwood+debattista_2009}.  This
  kind of behaviour is incompatible with any even approximate
  convergence at this value of $N$, but it does not exclude
  convergence at significantly larger values of $N$.  Indeed the
  apparent convergence at larger $N$ for this initial $b_0$ observed
  in Fig.~ \ref{FIG_I80_N} is associated with a very small dispersion
  in the results from realisation to realisation at larger $N$. As
we have discussed in our case this highly fluctuating behaviour occurs
in the region which appears to be stable for spherically symmetric
equilibria, but very close to the unstable region. It is however also
possible that it could be a marginally unstable region, for which a
recent analysis of a long-range toy model \citep{barre_etal_2016} has
shown that there can be a strong sensitivity of the evolution, and the
final state, to the initial conditions \citep{barre_etal_2016}.
In either case what we observe is presumably the result of the
competition between two different modes of instability around
stationary solutions which are so close to one another that their
 finite $N$ samplings at the given $N$ can overlap, while at larger
$N$ this will not be the case and a single mode of instability will
dominate.

\section{Discussion and conclusions} 

The main motivation for this study is to understand better whether the
results obtained with large N-body simulations of self-gravitating
systems represent faithfully the evolution, up to some time, of a
given collisionless system.  Let us consider now carefully what our
study allows us to conclude in this respect.

In a suite of N-body simulations at different $N$ of a simple class of
initial conditions we have shown that the macroscopic evolution
manifests two very different $N$ dependences.  On the one hand, in our
simulaions at smaller $N$ we have been able to observe that there is
an $N$ dependence coming from two-body collisional effects.  This is
quite straightforward to identify in a numerical study as a clearly
$N$ dependent drift of appropriately chosen macroscopic quantities, as
observed previously in similar systems
(e.g. \cite{theis+spurzem_1999}).  On the other hand there is a
distinct $N$ dependence associated with the presence of instabilities
in the collisionless dynamics.  This arises because the initial seeds
for the instability themselves depend on $N$. This dependence on $N$
is in practice much more difficult to find, as it manifests itself
only in a very weak dependence of the time of triggering of the
instability, and {\it not}, at sufficiently large $N$, in the
properties of the state to which the instability drives the
system. Indeed despite the fact that these initial conditions have
been studied fairly extensively in the literature, and that it might
seem evident to expect that there may be such a dependence on $N$,
this subtle dependence of the evolution on $N$ has apparently gone
unnoticed, other than in a study by \cite{boily+athanassoula_2006}.
These authors remarked, for cold initial conditions characterized by a
power law profile of density $\rho (r) \propto r^{-3/2}$, a very slow
convergence with $N$ of parameters characterizing the shape of the
final triaxial system obtained (similar to our
$\iota_{80}$). Specifically they concluded from a detailed study
analogous to ours in Section \ref{Ndependence-final}, that ``$N=10^5$
or larger is required for convergence of axial aspect ratios'' in
numerical experiments of this type. In light of our analysis it is
clear that this conclusion can be expected to depend strongly on the
initial condition. Further it does not apply to the temporal evolution
of the system. Indeed the latter should in fact be expected not to
converge at all as $N$ is extrapolated, and indeed it is not evident
even that final axis ratios should be expected to converge unless the
instability leading to breaking of symmetry occurs well after
virialisation. In the case analysed by \cite{boily+athanassoula_2006}
this appears not to be the case, and in general for very cold initial
conditions to reach such an asymptotic regime in a numerical study
would appear to be a considerable challenge.
 
We emphasise again that while the first type of $N$ dependence is
indicative of collisional effects, and therefore of effects which are
absent in the collisionless limit which it is the goal to simulate,
the second type of $N$ dependence is due to an instability which
exists in the collisionless limit --- indeed the radial orbit
instability is an instability of stationary solutions of the
collisionless Boltzmann equation. Thus the $N$ dependence observed in
the convergence study we have performed does not mean that the N-body
system does not approximate well the behaviour of a collisionless
self-gravitating system. It just means that it does not follow the
collisionless evolution of the exactly spherically symmetric initial
condition to which the $N \rightarrow \infty$ limit considered
converges. Instead when the instability develops, the N-body system is
following evolution from an initial condition with a finite
perturbation away from the exact spherical symmetry of the model
initial condition. To test numerically whether this evolution is
indeed collisionless one would need to perform a different large $N$
extrapolation of an N-body initial condition at some given $N=N_0$, in
which $N$ is varied keeping the fluctuations away from spherical
symmetry at $N_0$ fixed, i.e., one would need to resample at larger
and larger $N$ the density field of the $N=N_0$ simulation. The suite
of N-body simulations should then converge well at any finite
time. The difficulty with such a convergence study is that such a
resampling always introduces new fluctuations at smaller scales. One
does not know a priori whether these fluctuations may play some
significant role in the dynamics.  We plan to study this issue further
in future work.

Our study thus illustrates that, in performing a convergence study of
an N-body self-gravitating system to test whether it can be considered
to represent that of a collisionless system, there may be $N$
dependences which may be very much slower than those of two-body
effects in particular, and consequently much more difficult to
identify in a numerical study.  Indeed a generic instability sourced
by fluctuations, with amplitude of order $\frac{1}{\sqrt{N}}$ if
sampling is Poissonian, will lead to a dependence of the time it
develops at, which will grow in proportion to $\log N$ for an
exponential instability (assuming that numerical integration is
sufficiently accurate that numerical errors do not seed the
instability).  In the simple case we have chosen to study we have been
able to find such an $N$ dependence because it is associated with a
known instability, the role of which is easy to detect as it leads to
a change in the symmetry of the structure. In the generic case where
one does not know a priori that there are instabilities --- and do not
have evident tools for recognising that they may be present and
playing a central role in the dynamics --- one could thus easily miss
completely in a numerical convergence study that the results are in
fact $N$ dependent.

Although we have not analysed the specific case of cosmological
simulations we believe our results are very relevant to them. In this
respect the important point to note is that initial fluctuations of
cosmological simulations, although known theoretically at all scales,
are necessarily sampled in N-body simulations only at scales above the
initial grid spacing (see e.g. \cite{bertschinger_98}). When an
initially overdense region undergoes non-linear evolution the
collapsing region is qualitatively similar to that of the finite
system we have studied. The resulting collapsing structures typically
contain a very modest number of particles. Indeed for typical cold
dark matter initial conditions, such structure formation is
hierarchical, proceeding through the collapse of successively larger
structures starting from ones containing only a handful of
particles. The fluctuations associated with the finite number of
particles sampling any structure, which have no direct relation to
those of the continuum model's initial conditions, can then
potentially play a crucial part in the non-linear evolution of the
structure.  Indeed, in respect of the particular example we have
studied, we note that several works in the literature
\citep{huss_etal_1999, macmillan_etal_2006} find that the radial orbit
instability apparently plays a crucial role in the evolution of
cosmological halos, and even claim that it is this instability which
explains the apparent ``universality" of the profiles of cosmological
halos.  While existing convergence studies (see
e.g. \cite{power_etal_2002}) of such collapses appear to show that
these profiles are in fact $N$ independent, we note that they involve
the comparison of the ``final" state of the halos rather than the
detailed comparison of their temporal evolution. As we have seen here
it is the latter which shows up the subtle $N$ dependences which would
otherwise have escaped us. If radial orbit instability indeed plays a
central role in the evolution of simulated cosmological halos, it is
important to verify that the amplitude of the fluctuations driving the
instability are not determined by the finite number of particles used
to sample the initial conditions.  More generally, in the strongly
non-linear regime we should be very cautious in concluding that
results truly represent the relevant continuum limit on the basis of
convergence studies covering a very modest range of particle number.
These issues could be addressed further in a controlled manner
  by using initial conditions more closely representative of those of
  collapsing regions in N-body cosmological simulations, and, for
  example, by studying the effect of additional external forces
  mimicking the tidal forces inevitably present in such a setting.
Finally we note that our results provide also a strong motivation for
the use of different methods to study collisionless systems, and in
particular direct integration of the collisionless Boltzmann equation
as attempted by some recent studies \citep{yoshikawa_etal_2013,
  colombi_etal_2015, colombi+sousbie_2015}.

\bigskip 
Numerical simulations have been run on the Cineca Fermi cluster
(project VR-EXP HP10C4S98J), and on the HPC resources of The Institute
for Scientific Computing and Simulation financed by Region Ile de
France and the project Equip@Meso (reference ANR-10-EQPX- 29-01)
overseen by the French National Research Agency (ANR) as part of the
Investissements d'Avenir program. T.W. is supported by The Thailand
Research Fund (TRF) under contract number TRG5880036.

%\bibliographystyle{mn2e}
%\bibliography{bibliography}

\setlength{\bibhang}{2.0em}
\setlength\labelwidth{0.0em}

\end{document}